\documentclass{article}
\usepackage{amsmath,amsthm,amsfonts,amscd}
\newcommand{\p}{\partial}
\renewcommand{\=}{:=}

\theoremstyle{definition}

\theoremstyle{definition}
 
\begin{document}
\title{\bf Moufang loops and Lie algebras
\footnote{Presented at the 12th Colloquium ``Quantum Groups and Integrable Systems",
Prague, 12-14 June  2003}}
\date{}
\author{{\Large Eugen Paal}\\ \\
Department of Mathematics, Tallinn Technical University\\
Ehitajate tee 5, 19086 Tallinn, Estonia}
\maketitle \thispagestyle{empty}
\begin{abstract}
It is explicitly shown how the Lie algebras can be associated with the
analytic Moufang loops. The resulting Lie algebra commutation relations are
well known from the theory of alternative algebras and can be seen as a
preliminary step to \emph{quantum Moufang loops}.
\end{abstract}

\section{Moufang loops}

It is well known how the Lie algebras are connected with the Lie groups.
In the present paper, it is explicitly shown how the Lie algebras can be associated
with the analytic Moufang loops.

A Moufang loop \cite{RM,HP} is a quasigroup $G$ with the identity
element $e\in G$ and the Moufang identity
\[
(ag)(ha)=a(gh)a,\qquad a,g,h\in G.
\]
Here the multiplication is denoted by juxtaposition.
In general, the multiplication need not be associative: $gh\cdot a\neq g\cdot ha$.
Inverse element $g^{-1}$ of $g$ is defined by
\[
gg^{-1}=g^{-1}g=e.
\]
The left ($L$) and right ($R$) \emph{translations} are defined by
\[
gh=L_g h=R_h g,\qquad g,h\in G.
\]
Both translations are inverible mappings and
\[
L^{-1}_g=L_{g^{-1}},\qquad R^{-1}_g=R_{g^{-1}}
\]

\section{Analytic Moufang loops and infinitesimal Moufang translations}

Following the concept of the Lie group, the notion of an  analytic
Moufang loop can be easily formulated.

A Moufang loop $G$ is said \cite{Mal} to be \emph{analytic} if $G$ is also a real
analytic manifold and main operations - multiplication and the inversion map
$g\mapsto g^{-1}$ - are analytic mappings.

Let $T_e(G)$ denote the tangent space of $G$ at the unit $e$.
For $x\in T_e(G)$, infinitesimal Moufang translations are
defined as vector fields on $G$ as follows:
\begin{align*}
L_x&\=L_x(g)\=(dL_g)_ex\quad\,\in T_g(G),\\
R_x&\=R_x(g)\=(dR_g)_ex\quad\in T_g(G).
\end{align*}

Let the local coordinates of $g$ from the vicinity of $e\in G$ be
denoted by $g^{i}$ ($i=1,\dots,r\=\dim G$).
Define the auxiliary functions
\[
L^{i}_j(g)\=\frac{\p(gh)^{i}}{\p h^{j}}{\Big|}_{h=e},\quad
R^{i}_j(g)\=\frac{\p(hg)^{i}}{\p h^{j}}{\Big|}_{h=e}
\]
Matrices $(L^{i}_j)$ and $(R^{i}_j)$ are invertible.

Let $b_j\=\frac{\p}{\p g^{j}}|_e$ $(j=1,\dots,r)$ be a base in $T_e(G)$.
Then both
\begin{align*}
L_j&\=(dL_g)_eb_j=L^{i}_j(g)\frac{\p}{\p g^{i}}\quad\,\in T_g(G),\\
R_j&\=(dR_g)_eb_j=R^{i}_j(g)\frac{\p}{\p g^{i}}\quad\in T_g(G),
\end{align*}
give base at $T_g(G)$.
Thus one has two preferred base fields on $G$.
When writing $T_e(G)\ni x=x^{j}b_j$, one can easily see that
\begin{align*}
L_x&\=L_x(g)=x^{j}L^{i}_j(g)\frac{\p}{\p g^{i}}\quad\,\in T_g(G),\\
R_x&\=R_x(g)=x^{j}R^{i}_j(g)\frac{\p}{\p g^{i}}\quad\in T_g(G).
\end{align*}

\section{Structure constants and tangent Mal'tsev algebra}

As in the case of the Lie groups, the structure constants $c^{i}_{jk}$
of an analytic Moufang loop are defined by
\[
c^{i}_{jk}\=\frac{\p^{2}(ghg^{-1}h^{-1})^{i}}{\p g^{j}\p h^{k}}\Big|_{g=h=e}=-c^{i}_{kj},
\qquad i,j,k=1,\dots,r.
\]
For any $x,y\in T_e(G)$, their 
product $[x,y]\in T_e(G)$ is
defined in component form by
\[
[x,y]^{i}\=c^{i}_{jk}x^{j}y^{k}=-[y,x]^{i},\qquad i=1,\dots,r.
\]
The tangent space $T_e(G)$ being equipped with such an anti-commutative
multiplication is called the \emph{tangent algebra} of the analytic
Moufang loop $G$.

The tangent algebra of $G$ need not be a Lie algebra. There may exist a triple
$x,y,z\in T_e(G)$ which does not satisfy the Jacobi identity:
\[
J(x,y,z)\=[x,[y,z]]+[y,[z,x]]+[z,[x,y]]\neq0.
\]
Instead, for any $x,y,z\in T_e(G)$ one has a more general \emph{Mal'tsev identity}
\[
[J(x,y,z),x]=J(x,y,[x,z]).
\]
Anti-commutative algebras with this identity are called the \emph{Mal'tsev algebras}.

\section{Generalized Maurer-Cartan equations}

Denote as above $L_x\=L_x(g)$ and $R_x\=R_x(g)$ for all $x\in T_e(G)$.

It is well known that the infinitesimal translations of a \emph{Lie group }
obey the \emph{Maurer-Cartan equations}
\[
[L_x,L_y]-L_{[x,y]}=[L_x,R_y]=[R_x,R_y]+R_{[x,y]}=0.
\]
It turns out that for a non-associative analytic Moufang loop these equations
are violated minimally. The algebra of infinitesimal Moufang translations reads
as \emph{generalized Maurer-Cartan equations}:
\[
[L_x,L_y]-L_{[x,y]}=-2[L_x,R_y]=[R_x,R_y]+R_{[x,y]}
\]
We outline a way of closing of this algebra (generalized Maurer-Cartan
equations), which in fact means construction of a \emph{finite} dimensional Lie
algebra generated by infinitesimal Moufang translations.

Start by rewriting the generalized Maurer-Cartan equations as follows:
\begin{align}
[L_x,L_y]&=\,\,2Y(x;y)+\frac{1}{3}L_{[x,y]}+\frac{2}{3}R_{[x,y]}\\
[L_x,R_y]&=-Y(x;y)+\frac{1}{3}L_{[x,y]}-\frac{1}{3}R_{[x,y]}\\
[R_x,R_y]&=\,\,2Y(x;y)-\frac{2}{3}L_{[x,y]}-\frac{1}{3}R_{[x,y]}
\end{align}
Here (1) or (2) or (3) can be assumed as a definition of the \emph{Yamagutian} $Y$.
It can be shown that
\begin{gather}
Y(x;y)+Y(y;x)=0,\\
Y([x,y];z)+Y([y,z];x)+Y([z,x];y)=0.
\end{gather}
The constraints (4) trivially descend from the anti-commutativity of the commutator
bracketing, but the proof of (5) needs certain effort. Further, it turns out
that the following \emph{reductivity} conditions hold:
\begin{equation}
6[Y(x;y),L_z]=L_{[x,y,z]},\qquad
6[Y(x;y),R_z]=R_{[x,y,z]}
\end{equation}
where the trilinear \emph{Yamaguti brackets} $[\cdot,\cdot,\cdot]$ are defined
\cite{Yam} in $T_e(G)$ by
\[
[x,y,z]\=[x,[y,z]]-[y,[x,z]]+[[x,y],z].
\]
Finally, the Yamagutian obeys the Lie algebra
\begin{equation}
6[Y(x;y),Y(z;w)]=Y([x,y,z];w)+Y(z;[x,y,w]).
\end{equation}
The Lie algebra commutation relations (1)-(7) were proved in \cite{Paal}.
Dimension of this Lie algebra does not exceed
$2r+r(r-1)/2$. The Jacobi identities are guaranteed
by the defining identities of the Lie \cite{Loos} and general Lie \cite{Yam}
\emph{triple systems} associated with the tangent Mal'tsev algebra $T_e(G)$ of $G$.

The commutation relations of form (1)-(7) are well-known from the theory of
alternative algebras \cite{Schafer} and can be seen as a preliminary step
to construct \emph{quantum Moufang loops} (QML). QML is a
deformation of universal enveloping algebra of the Lie algebra (1)-(7).


The paper was in part supported by the Estonian Science Foundation, Grant 5634.

\end{document}